\documentclass{emulateapj}

\submitted{Received 2004 December 23; accepted 2005 January 20}

\lefthead{Lee et al.}
\righthead{Super He-rich Population and the Origin of EHB Stars}

\begin{document}

\title{SUPER HELIUM-RICH POPULATION AND THE ORIGIN OF EXTREME HORIZONTAL-BRANCH STARS IN GLOBULAR CLUSTERS}

\author{
Young-Wook Lee\altaffilmark{1},
Seok-Joo Joo\altaffilmark{1},
Sang-Il Han\altaffilmark{1},
Chul Chung\altaffilmark{1},
Chang H. Ree\altaffilmark{1},
Young-Jong Sohn\altaffilmark{1},
Yong-Cheol~Kim\altaffilmark{1},
Suk-Jin Yoon\altaffilmark{2},
Sukyoung K. Yi\altaffilmark{2}, and
Pierre Demarque\altaffilmark{3}
}

\altaffiltext{1}{Center for Space Astrophysics and Department of Astronomy, Yonsei University, Seoul 120-749, Korea; ywlee@csa.yonsei.ac.kr}
\altaffiltext{2}{Department of Physics, University of Oxford, Oxford OX1 3RH, UK}
\altaffiltext{3}{Department of Astronomy, Yale University, New Haven, CT 06520-8101, USA}

\begin{abstract}
Recent observations for the color-magnitude diagrams (CMDs) of the massive globular cluster $\omega$~Centauri have shown that it has a striking double main sequence (MS), with a minority population of bluer and fainter MS well separated from a majority population of MS stars. Here we confirm, with the most up-to-date $Y^2$ isochrones, that this special feature can only be reproduced by assuming a large variation (Î$\Delta$Y = 0.15) of primordial helium abundance among several distinct populations in this cluster. We further show that the same helium enhancement required for this special feature on the MS can by itself reproduce the extreme horizontal-branch (HB) stars observed in $\omega$~Cen, which are hotter than normal HB stars. Similarly, the complex features on the HBs of other globular clusters, such as NGC~2808, are explained by large internal variations of helium abundance. Supporting evidence for the helium-rich population is also provided by the far-UV (FUV) observations of extreme HB stars in these clusters, where the enhancement of helium can naturally explain the observed fainter FUV luminosity for these stars. The presence of super helium-rich populations in some globular clusters suggests that the third parameter, other than metallicity and age, also influences CMD morphology of these clusters.
\end{abstract}


\section{INTRODUCTION}
Discovery of the discrete multiple stellar populations in the most massive globular cluster $\omega$~Centauri (Lee et al. 1999) has stimulated more detailed follow up observations for this unique cluster, which led to the subsequent new discoveries. The most surprising feature found from these observations is a presence of double main-sequence (MS), with a minority population of bluer and fainter MS clearly separated from a majority population of redder and brighter MS (Anderson et al. 2002; Bedin et al. 2004). Recent observations have also confirmed the discrete nature of the red-giant-branches (RGBs) and a substantial population of extreme horizontal-branch (HB) stars, which are hotter than and well separated from the majority population of normal HB stars (Ferraro et al. 2004; Sollima et al. 2005). More recent analysis, based on the old Revised Yale isochrones (Green et al. 1987), has suggested that a large range of primordial helium abundance variation is needed to reproduce the double MS observed in $\omega$~Cen (Norris 2004). Spectroscopic measurement of metal abundances for the MS stars in this cluster (Piotto et al. 2005) has further reinforced the requirement of a large helium abundance for the bluer and fainter MS. Although more work is needed for a specific chemical evolution model devoted for $\omega$~Cen to understand the origin of this unusually strong helium enrichment, this would be a consequence of more than one epoch of star formation in some globular clusters, where the minority populations of later generations of stars would have been enriched in helium by the previous generations of stars (D'Antona \& Caloi 2004; Norris 2004; Piotto et al. 2005).

The purpose of this Letter is to show that the large variation of helium abundance in $\omega$~Cen can not only explain the special feature on the MS as has been suggested, but also naturally reproduce hot extreme HB stars observed in the same cluster. We further demonstrate that the complex features on the HBs of other globular clusters, such as NGC~2808, are explained by similarly large variations of helium abundance. This, together with other recent results described above, would stimulate more detailed modelling and observations for these clusters and also for other globular clusters with peculiar features on their HBs.
\begin{figure}
\epsscale{1.0}
\plotone{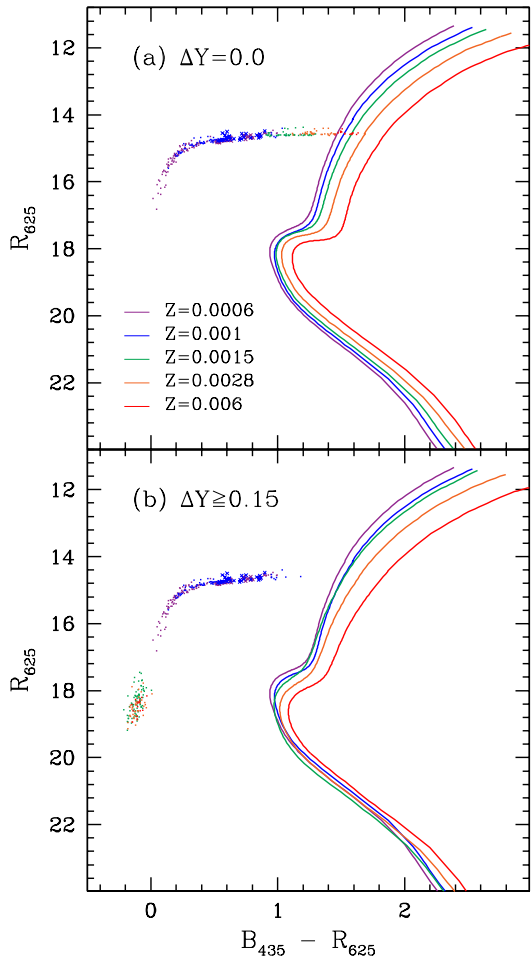}
\caption{Population models (synthetic HBs and new $Y^2$ isochrones) for $\omega$~Cen. Panel (a) is for the case that all five populations in $\omega$~Cen have the same helium abundance despite their different metallicities, while panel (b) is for the case that the helium abundances for the metal-rich populations are significantly enhanced as in Table 1. Only the case of $\Delta$Y $\geqq$ 0.15 can reproduce the observed features on the MS and HB, simultaneously.\label{fig1}}
\end{figure}

\section{POPULATION MODELS}
In order to investigate the effect of helium abundance on the observed features in the color-magnitude diagram (CMD), we have first calculated new sets of Yonsei-Yale ($Y^2$) isochrones (Kim et al. 2002) with the most up-to-date input physics, under different assumptions regarding the primordial helium abundance. Based on these isochrones and HB evolutionary tracks of Sweigart (1987) extrapolated to Y $\sim$ 0.4, population synthesis models are constructed following the techniques developed by Lee et al. (1990) and Park \& Lee (1997). First, in Figure 1a, we have presented the models for the case that all five populations (Sollima et al. 2005) in $\omega$~Cen have the same helium abundance despite their different metallicities. Then, the models in Figure 1b are constructed with the assumption that the helium abundances for the three most metal-rich populations in this cluster are significantly enhanced. In practice, the values of helium abundance and age are adjusted until the best matches between the models and the observed CMDs are obtained, while those for the metallicity are mostly fixed by the RGBs. Table 1 lists the input parameters used in our best model simulations\footnotemark[4]. Figure 2 is our synthetic CMDs for the MS and sub-giant-branch (SGB) parts only, constructed based on the isochrones in Figure 1b and other parameters listed in Table 1. These model CMDs are specifically constructed with the band passes, photometric errors, and total numbers of stars comparable to the observed CMDs of Bedin et al. (2004; see their Figure 1), so that they can be directly compared with the observed CMDs. As is clear from Figures 1 \& 2, we have, using the most up-to-date stellar evolution theory, confirmed the conclusion from previous works; a large variation in helium abundance is indeed needed to reproduce the observed features on the MS to RGB, including the double MS, where the minority population of bluer and fainter sequence is both more metal-rich and super helium-rich ($\Delta$Y = 0.15) compared to the majority population of metal-poor redder MS. Because of the difference in core temperature, helium-rich stars, in general, are brighter and bluer for given mass. Since they evolve faster than helium-poor stars, helium-rich stars would have smaller masses for given age, and therefore helium-rich MS appears both bluer and fainter than helium-poor sequence on the isochrone.
\footnotetext[4]{The mean values of the mass-loss on the RGB were estimated with the Reimers (1977) parameter $\eta$ = 0.46, which was obtained by fitting the model HB to the most metal-poor HB of $\omega$~Cen at 13 Gyr. This value of $\eta$ was used in all synthetic HB models in this paper. The mass dispersion on the HB required to fit the observed HBs are $\sim$ 0.01 $M_\odot$ for the three most metal-rich populations, and $\sim$ 0.02 $M_\odot$ for the two most metal-poor populations, respectively.}
\begin{figure}
\epsscale{1.0}
\plotone{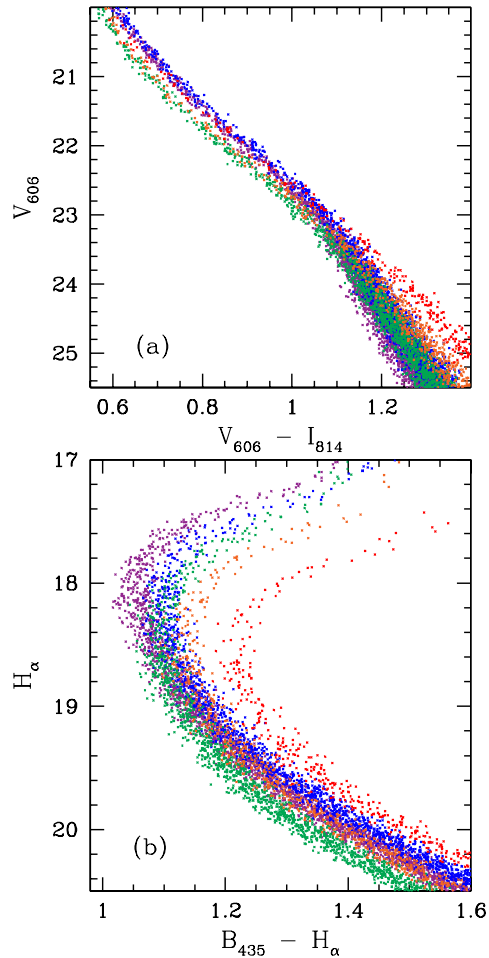}
\caption{Synthetic CMDs for the MS and SGB region of $\omega$~Cen. The reader is encouraged to compare these models with the observations by Bedin et al. (2004; see their Fig. 1) to see how well the models reproduce the observed features.\label{fig2}}
\end{figure}
In Figure 1, we have also presented corresponding synthetic HB models, which should be compared with the observed CMD by Ferraro et al. (2004; see their Figure 2). From models in Figure 1a, we can see that the HB morphology generally gets redder with increasing metallicity, and the models fail to reproduce the extreme HB population. Note, however, that models in Figure 1b naturally reproduce the extreme HB stars observed in $\omega$~Cen when adopting the same helium enhancements needed to reproduce the unique features on the MS. For the same age and metal abundance, helium-rich HB stars have smaller total masses than do less helium-rich stars, which shifts their positions to the blue (Lee et al. 1994). Therefore, the fact that three most metal-rich populations, including progeny of the bluer and fainter MS, all piled up on the extremely blue HB in Figure 1b is a consequence of the large difference in helium abundance that easily overcomes the metallicity effect. These extreme HB stars have very thin envelope outside the helium core, and hence the surface temperature becomes extremely hot. In our model calculations, nothing other than standard Reimers (1977) empirical relation was employed to estimate the amount of mass-loss on the RGB as a function of input parameters adopted (see Table 1), and therefore the presence of extreme HB stars is solely the effect of helium enhancement.
\begin{table}
\caption{Input parameters in our best simulation of $\omega$~Cen}
\begin{center}
\begin{tabular}{cccccc}
\tableline
\tableline
Population & $Z$\tablenotemark{a} & $Y$ & Age(Gyr) & Mass-loss($M_\odot$)\tablenotemark{b} & Fraction\tablenotemark{c} \\
\tableline
1 & 0.0006 & 0.231 & 13   & 0.168 & 0.42 \\
2 & 0.001  & 0.232 & 13   & 0.178 & 0.27 \\
3 & 0.0015 & 0.38  & 12   & 0.172 & 0.17 \\
4 & 0.0028 & 0.40  & 11.5 & 0.177 & 0.08 \\
5 & 0.006  & 0.42  & 11.5 & 0.201 & 0.05 \\
\tableline
\tableline
\end{tabular}
\end{center}
\tablenotetext{a}{[$\alpha$/Fe] = 0.3}
\tablenotetext{b}{Mean mass-loss on the RGB for $\eta$ = 0.46.}
\tablenotetext{c}{Population ratios adopted from Sollima et al. (2005).}
\end{table}
The good agreements between the models and the observations for the appearance and population ratio of extreme HB stars in $\omega$~Cen suggest that the third parameter that controls HB morphology, in addition to metallicity and age, might be helium abundance (D'Antona et al. 2002; D'Antona \& Caloi 2004). In order to test this hypothesis, in Figure 3, we have compared our model synthetic CMD with the observed CMD for NGC~2808, which represents the most extreme case of the peculiar HB morphology with two groups of extreme HB stars in addition to the well-known bimodal HB (Bedin et al. 2000). In our models in Figure 3b, helium abundance varies from Y = 0.23 (red HB) to Y = 0.43 (bluest extreme HB) among four distinct populations postulated in this model cluster. Note that the mean helium abundance, when considering the population ratio, becomes Y = 0.288, which is not inconsistent with the most recent estimation using the R-method (Salaris et al. 2004), to within the errors. Unlike $\omega$~Cen, no internal metallicity dispersion was employed as the RGB of NGC~2808 has rather narrow color spread. As is clear from Figure 3, the models well reproduce the observed peculiar features on the HB. On the other hand, the internal helium abundance variation has relatively weak effect on the RGB and MS, and hence it would not be detected unless one has extremely high precision data. Note, however, that the variation of helium abundance predicted from the HB morphology of NGC~2808 is comparable to that suggested in $\omega$~Cen, while we expect no or negligible metallicity spread. Therefore, we predict either the split or broadened feature on the MS also in NGC~2808. High precision $Hubble~Space~Telescope$ ($HST$) photometry is highly anticipated in this regard.
\begin{figure}
\epsscale{1.0}
\plotone{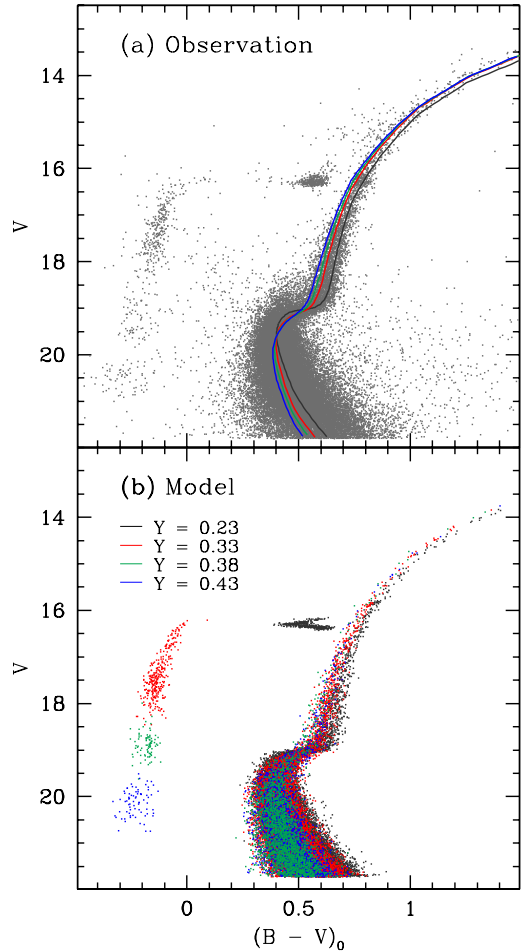}
\caption{Comparison of models with the observed CMD (Bedin et al. 2000) for NGC~2808. The synthetic CMD is constructed with the assumption that four distinct populations in this cluster have different helium abundances but same metallicity (Z = 0.0014) and age (10.1 Gyr).\label{fig3}}
\end{figure}
Independent test for the helium-rich population is provided from the FUV photometry of extreme HB stars, as the zero-age HB (ZAHB) locus is very sensitive to helium abundance. In general, helium-rich HB stars are brighter, however, this trend is reversed when the surface temperature reaches $\sim$ 19,000K. This is because extremely hot HB stars have very thin envelope, with almost negligible energy output from the hydrogen-burning shell. The total energy output is then mostly sustained by the helium-burning core, and since helium-rich stars have smaller core mass, they have lower surface luminosity (Sweigart 1987). According to our models, the extreme HB stars, therefore, would be fainter than the redder HB stars as a result of difference in helium abundance. In Figure 4, we have compared $HST/STIS$ UV CMD of extreme HB and blue HB stars in NGC~2808 (Brown et al. 2001) with the models constructed in the same band passes based on the ZAHB loci and post-ZAHB evolutionary tracks of Sweigart (1987). Figure 4a is for the case that all stars have the same helium abundance of Y = 0.23, where the extreme HB stars appear clearly fainter than the redder HB stars and spread well below the ZAHB locus. Figure 4b demonstrates that a reasonable agreement is obtained when we adopt a large range of helium abundance as in our models in Figure 3. The same trend is also observed in $\omega$~Cen (Whitney et al. 1994; see their Fig. 9).

\section{DISCUSSION}
The presence of extreme HB or other peculiar features on the HB (i.e., the third parameter problem), in principle, can be reproduced by internal variation of parameters, such as mass-loss on the RGB, helium abundance, and core rotation. The evidence presented above strongly suggests, however, that the third parameter problem of the HB morphology is largely, if not all, due to the helium abundance variations among two or more stellar populations in some globular clusters. This in turn suggests that whenever the relative age is estimated from the HB morphology (Lee et al. 1994; Rey et al. 2001), better result would be obtained by ignoring extreme HB stars or similarly peculiar features on the HB. Fortunately, the effect of helium abundance on the MS and RGB is relatively small, and in most globular clusters with extreme HB stars, the expected helium-rich populations are minority in terms of population ratio. Therefore, the age dating from MS and RGB is likely to be less affected by the minority population of helium rich stars in these clusters. Nevertheless, when one infers relative age or other physical parameters from the integrated-light colors and spectra of globular clusters, possible contamination from helium-rich populations should still be carefully considered, as they would make the observed colors bluer, especially in UV.
\begin{figure}
\epsscale{1.0}
\plotone{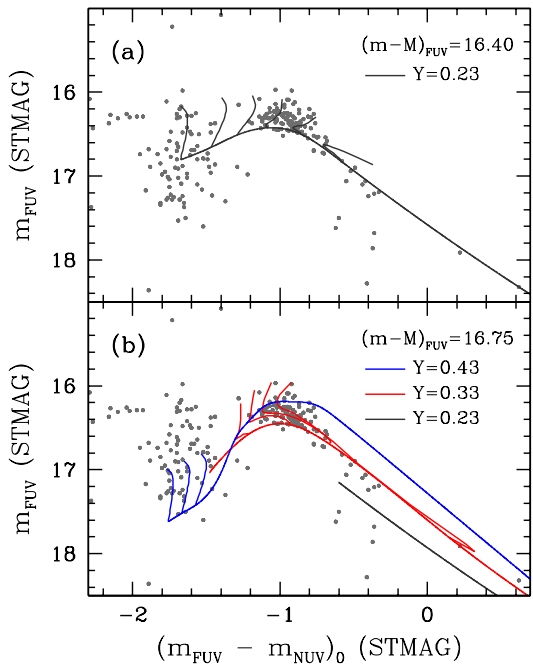}
\caption{Comparison of $HST/STIS$ UV CMD of extreme HB and blue HB stars in NGC~2808 (Brown et al. 2001) with the model predictions. Panel (a) is for the case that all stars have the same helium abundance of Y = 0.23, while panel (b) is for the case of a large range of helium abundance as in our models in Figure 3.\label{fig4}}
\end{figure}
It appears unlikely, however, that this unusual helium enrichment postulated in peculiar globular clusters would also occur in galactic scales. The helium enrichment parameter, $\Delta$Y/$\Delta$Z, observed and expected in galaxies is about 2 (e.g., Jimenez et al. 2003), while the values required in the above globular clusters are more than 90. The metallicity distribution function (MDF) of $\omega$~Cen differs significantly from those of the Galactic bulge and elliptical galaxies, in that there are many more metal-poor stars than metal-rich stars in $\omega$~Cen (Lee et al. 1999; Sollima et al. 2005), while the opposite is the case in the Galactic bulge and elliptical galaxies (e.g., Rich 1988). While we need more detailed chemical evolution studies to understand the origin of this unusual helium enrichment in $\omega$~Cen, the difference in MDF suggests that the enrichment would be much more efficient in the systems like $\omega$~Cen. It appears, therefore, the extreme HB stars observed in some globular clusters are most likely a local effect. Further observations and modelling for the globular clusters with peculiar features on their HBs will undoubtedly help to establish the complex star formation histories and accompanying chemical enrichments in these stellar systems, along with the possible connection between these globular clusters and the relicts of the Galaxy building blocks (Ibata et al. 1994; Lee et al. 1999).
\acknowledgments{Support for this work was provided by the creative research initiative program of the Korean ministry of science \& technology, for which we are grateful. P.D. acknowledges partial support from NASA grant NAG5-13299.}


\begin{references}
\reference{}Anderson, J. 2002, in ASP Conf. Ser. Vol. 265, $\omega$~Centauri: A Unique Window into Astrophysics, eds F. van Leeuwen, J. D. Hughes, \& G. Piotto. (San Francisco: ASP), 87
\reference{}Bedin, L. R., Piotto, G., Anderson, J., Cassisi, S., King, I. R., Momany, Y., \& Carraro, G. 2004, \apj, 605, L125
\reference{}Bedin, L. R., Piotto, G., Zoccali, M., Stetson, P. B., Saviane, I., Cassisi, S., \& Bono, G. 2000, \aap, 363, 159
\reference{}Brown, T. M., Sweigart, A. V., Lanz, T., Landsman, W. B., \& Hubeny, I. 2001, \apj, 562, 368
\reference{}D'Antona, F., \& Caloi, V. 2004, \apj, 611, 871
\reference{}D'Antona, F., Caloi, V., Montalban, J., Ventura, P., \& Gratton, R. 2002, \aap, 395, 69
\reference{}Ferraro, F. R., Sollima, A., Pancino, E., Bellazzini, M., Straniero, O., Origlia, L., \& Cool, A. M. 2004, \apj, 603, L81
\reference{}Green, E. M., Demarque, P., \& King, C. R. 1987, The Revised Yale Isochrones and Luminosity Functions (New Haven: Yale Univ. Obs.)
\reference{}Ibata, R. A., Gilmore, G., \& Irwin, M. J. 1994, Nature, 370, 194
\reference{}Jimenez, R., Flynn, C., MacDonald, J., \& Gibson, B. K. 2003, Science, 299, 1552
\reference{}Kim, Y.-C., Demarque, P., Yi, S. K., \& Alexander, D. R. 2002, \apjs, 143, 499
\reference{}Lee, Y.-W., Joo, J.-M., Sohn, Y.-J., Rey, S.-C., Lee, H.-c., \& Walker, A. R. 1999, Nature, 402, 55
\reference{}Lee, Y.-W., Demarque, P., \& Zinn, R. 1990, \apj, 350, 155
\reference{}Lee, Y.-W., Demarque, P., \& Zinn, R. 1994, \apj, 423, 248
\reference{}Norris, J. E. 2004, \apj, 612, L25
\reference{}Park, J.-H., \& Lee, Y.-W. 1997, \apj, 476, 28
\reference{}Piotto, G., et al. 2005, \apj, in press
\reference{}Reimers, D. 1977, \aap, 57, 395
\reference{}Rey, S.-C., Yoon, S.-J., Lee, Y.-W. Chaboyer, B., \& Sarajedini, A. 2001, \aj, 122, 3219
\reference{}Rich, R. M. 1988, \aj, 95, 828
\reference{}Salaris, M., Riello, M., Cassisi, S., \& Piotto, G. 2004, \aap, 420, 911
\reference{}Sollima, A., Ferraro, F. R., Pancino, E., \& Bellazzini, M. 2005, \mnras, in press
\reference{}Sweigart, A. 1987, \apjs, 65, 95
\reference{}Whitney, J. H., et al. 1994, \aj, 108, 1350
\end{references}
\end{document}